# Bulk Photovoltaic Effects in Helimagnets

Chunmei Zhang[1], Hanqi Pi[2], Jian Zhou[3,*]

[1]School of Physics, Northwest University, Xi'an 710127, China
[2]Beijing National Laboratory for Condensed Matter Physics and Institute of Physics, Chinese Academy of Sciences, Beijing 100190, China
[3]Center for Alloy Innovation and Design, State Key Laboratory for Mechanical Behavior of Materials, Xi'an Jiaotong University, Xi'an 710049, China

[*]Email: jianzhou@xjtu.edu.cn

Abstract

The bulk photovoltaic (BPV) effect that converts light into electric current is highly sensitive to the system symmetry and its electronic Bloch wavefunction. To create a sizable net electric current, it is necessary to break the centrosymmetry $\mathcal{P}$ in its host material. While prior studies mainly focus on $\mathcal{P}$-broken nonmagnetic (time reversal $\mathcal{T}$-reserved) and collinear antiferromagnetic $\mathcal{PT}$ systems, here we adopt magnetic group theory to scrutinize how helical spin polarization breaks $\mathcal{P}$ that is otherwise kept in its crystalline geometry. Such spin helix widely exists in intrinsic multiferroics and magnon excitation in collinear magnets. We demonstrate that magnetic symmetries are determined by spin spiral configurations, helical propagation wavevectors, magnetic chirality, and spin winding angles at the home site, offering effective and versatile manipulation of BPV generation in helimagnets. This is illustrated by tight-binding model calculations. Furthermore, we apply our theory to monolayer NiI$_2$, a realistic showcase for helimagnetic material. Our first-principles calculations show that it could host observable direction-dependent BPV current, which could serve as a photoelectric probe to track the subtle magnetic configurations and potential multiferroic nature. This work reveals the fundamental mechanism of the interplay between spin spiral order and nonlinear optics process.

*Introduction.* Bulk photovoltaic (BPV) effect that produces electric current in a single material with broken centrosymmetry ($\mathcal{P}$) under homogeneous light illumination [1-4], could avoid the complexity of constructing heterojunctions and engineering defects in conventional photoelectric devices. Among various mechanisms, shift current and ballistic current [5,6] are widely studied and have attracted much attention. The shift current originates from coherent wavefunction center evolution between the valence and conduction bands. The ballistic current arises from asymmetric carrier generation at $\boldsymbol{k}$ and $-\boldsymbol{k}$ such as impurities or electron-phonon coupling. Intrinsic ballistic current is also known as injection current, requiring either circularly polarized light (CPL) or magnetic materials to break time-reversal ($\mathcal{T}$) symmetry [7]. In addition, kinetic processes (radiative recombination and scattering current) [8,9] also contribute to the BPV effect. In nonmagnetic $\mathcal{P}$-broken systems, shift and injection current are induced by linearly and circularly polarized light, respectively.

Recent attentions have been focusing on collinear antiferromagnetic materials with $\mathcal{PT}$ symmetry [10-17]. It has been shown that circular and linear light could trigger shift and injection current, respectively. As such, four different intrinsic BPV currents could emerge in magnetic systems, i.e., linearly polarized light (LPL) could trigger normal shift current (NSC) and magnetic injection current (MIC), while CPL may generate normal injection current and magnetic shift current [10,18-20]. All of them are rooted in topological phases of electron wavefunctions, such as shift vectors, Berry curvature, and quantum metrics. While most prior works focus on nonmagnetic $\mathcal{P}$-broken and collinear antimagnetic ($\mathcal{PT}$-preserved) systems, in this Letter, we scrutinize a more general case, i.e., how spiral spin patterns generate direction-dependent BPV current in helimagnetic materials, which belong to centrosymmetric crystalline point group when spin is not considered. Note that the collective spin spiral pattern, which largely appears in intrinsic multiferroic materials and magnon excitation of collinear magnets, offers ultrafast dynamics and low energy dissipation in information storage and manipulation processes [21]. Until now, the spiral spin determined magnetic symmetries and their control of BPV generation remain unexplored.

We conduct our theory using two prototypical helimagnetic patterns, namely, proper screw and (in-plane) cycloid structures. They refer to the spin helical plane perpendicular and parallel to the magnetic propagation vector $\boldsymbol{q}$, respectively. It has been shown that the spin helix induces electric polarization according to $\boldsymbol{P}_{ij} = A\boldsymbol{e}_{ij} \times (\boldsymbol{S}_i \times \boldsymbol{S}_j) + B[(\boldsymbol{e}_{ij} \cdot \boldsymbol{S}_i)\boldsymbol{S}_i - (\boldsymbol{e}_{ij} \cdot \boldsymbol{S}_j)\boldsymbol{S}_j]$, where $A$ and $B$ are the coupling parameters arising from spin-orbit coupling (SOC), and $\boldsymbol{e}_{ij}$ is a unit vector pointing from spin sites $\boldsymbol{S}_j$ to $\boldsymbol{S}_i$ [22-24]. The first term describes the cycloid pattern from the inverse Dzyaloshinskii-Moriya interaction, and the second term gives rise to the polarization in the proper screw texture [25]. Hence, helimagnetic configurations host finite $\boldsymbol{P}$ that are strongly coupled to their spin texture, giving a type-II multiferroic character. We adopt a two-dimensional (2D) trigonal lattice and derive a general theory to show that their magnetic groups are simultaneously determined by the specific spin configuration, helimagnetic wavevector $q$, and the winding angle at the home site. The BPV components are investigated by tight-binding model calculations, revealing clear symmetry constrained directional dependence. We further use a realistic material, monolayer $NiI_2$, as a showcase to illustrate this concept. The $NiI_2$ belongs to the transition-metal dichlodide family [25-31], whose thin film and bulk phases have been revealed to be helimagnetic [32]. In particular, the multiferroic nature of monolayer $NiI_2$ is under extensive investigations recently [33-36]. We show a unidirectional BPV of ~50 $\mu A/V^2$, offering observable magnitude in experiments.

*Model and symmetry analysis.* We denote a magnetic propagation wavevector $q$ to describe the helimagnetic pattern. In general, the local magnetic moment takes the form

$$\boldsymbol{M}_i = M(\cos\varphi_i \sin\theta_i, \sin\varphi_i \sin\theta_i, \cos\theta_i) \tag{1}$$

Here, $\theta_i$ is the polar angle and $\varphi_i$ is the azimuthal angle of the local magnetic polarization vector at magnetic site $i$. $M$ is the magnetization magnitude, describing local exchange field. The proper screw helimagnetic configuration [Fig. 1(a)] is

$$\varphi_i = \frac{\pi}{2}, \theta_i = \theta_0 + \boldsymbol{q} \cdot \boldsymbol{r}_i, \tag{2}$$

and the cycloid pattern [Fig. 1(b)] gives

$$\theta_i = \frac{\pi}{2}, \varphi_i = \varphi_0 + \boldsymbol{q} \cdot \boldsymbol{r}_i, \tag{3}$$

Here, $\boldsymbol{r}_i$ is the position of the magnetic site with respect to lattice vectors, and $\varphi_0$ and $\theta_0$ are winding angles at the home site ($\boldsymbol{r}_i = \boldsymbol{0}$).

For simplicity, we assume that each unit cell has one magnetic site. The $\boldsymbol{q} = (q_1, q_2, q_3)$ is measured with respect to reciprocal lattice of the unit cell. For each index $j$ ($j = 1, 2, 3$), we assume that $q_j = \frac{n_j}{\ell_j} \in \left[-\frac{1}{2}, \frac{1}{2}\right]$ which could represent the entire Brillouin zone (BZ). Note that $q_j = 0$ and $\pm\frac{1}{2}$ refer to ferromagnetic and collinear antiferromagnetic configurations, respectively. We then apply group theory to analyze the symmetry in helimagnetic systems with general $q_j$. Without considering spin configuration, the system belongs to crystalline point group $\mathcal{G}$ that is centrosymmetric. The collinear ferromagnetic patterns belong to six magnetic point groups (MPGs), namely, $2'm'$, $m'm'm$, $2/m$, $4/m$, $4/mm'm'$, and $6/mm'm'$. In these cases, the magnetic unit cell remains the same as crystalline unit cell. However, the spiral spin pattern reduces such lattice symmetry with a larger magnetic supercell. Hence, they would in general belong to the type-IV Shubnikov group, derived from a subgroup of $\mathcal{G}$.

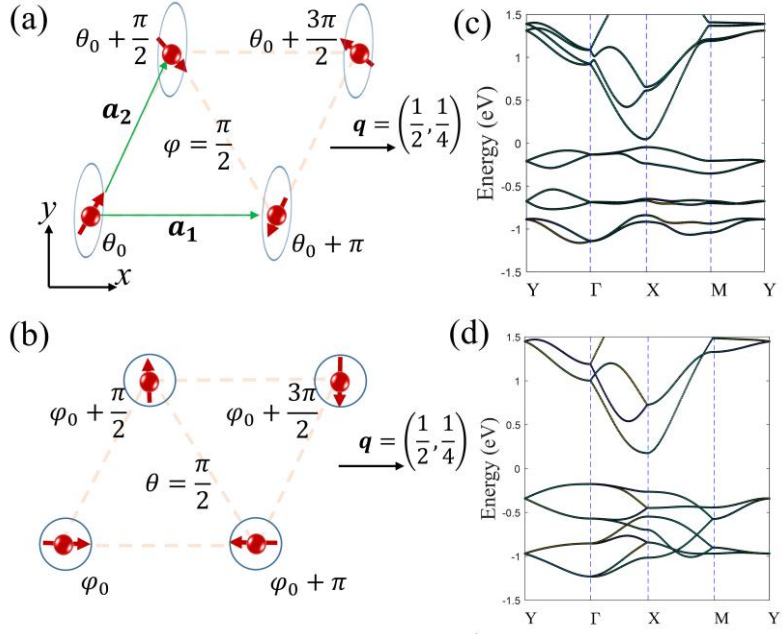

FIG. 1. Schematic plot of the helimagnetic states for the (a) proper screw and (b) cycloid patterns in a trigonal lattice, with $\boldsymbol{q} = \left(\frac{1}{2}, \frac{1}{4}\right)$. Their tight-binding band dispersion is plotted in (c) with $\theta_0 = \frac{\pi}{18}$ and (d) with $\varphi_0 = 0$, respectively. The hopping parameters are $V_{dd\sigma} = -0.4$, $V_{dd\pi} = 0.3$, and $V_{dd\delta} = 0.6$. The on-site $U$ values for $d_{xy}$, $d_{x^2-y^2}$, and $d_{z^2}$ are set to be 0, 0, and 4, respectively. The universal magnetic exchange magnitude is $M = 0.7$. The SOC parameter is $t_{\text{soc}} = 0.7$. All of them are in unit of eV.

We take a 2D trigonal lattice to conduct the symmetry analyses. The magnetic propagation wavevector is $\boldsymbol{q} = \left(\frac{n_1}{\ell_1}, \frac{n_2}{\ell_2}\right)$, $(\ell_1, \ell_2 \in \mathbb{N}, n_1, n_2 \in \mathbb{Z})$. Here, $-\frac{\ell_i}{2} \le n_i \le \frac{\ell_i}{2}$ $(i = 1,2)$, with $n_i = 0$ describing ferromagnetic alignment. The magnetic supercell is then $(\ell_1 \times \ell_2)$, and we denote a potential translation operator by $\boldsymbol{t}_{v_1,v_2} = (v_1, v_2)$, $(v_i \in \mathbb{Z}, |v_i| \le \frac{\ell_i}{2})$, measured with respect to the unit cell. Regardless of spin helix, its crystalline layer group is $P6/mmm$, corresponding to $\mathcal{G} = D_{6h}$. Under spin spiral configuration, both inversion symmetry and three-fold rotation no longer exist, so that the MPG will be derived by two-fold rotations and/or mirror reflections, namely, from point group of 222 and $mm2$ and their subgroups. We perform careful derivation for each operation, which can be found in Supplemental Material [37]. It is disclosed that both wavevector $\boldsymbol{q}$ and the winding angles at the home site ($\theta_0$ and $\varphi_0$) strongly influence the symmetry and BPV generation. A particular interesting situation lies when $\frac{n_1}{\ell_1} = \frac{2n_2}{\ell_2}$, which is highly symmetric than other cases, and has been proposed to appear in realistic materials (see below). The MPG and allowed BPV components (under LPL) are tabulated in Table I. Here, we find that the even and odd $\ell_2$ determines whether they belong to gray and black-white MPGs, respectively. In the former case, the MIC would totally vanish. One sees that in some cases, both NSC and MIC are symmetrically forbidden, even though the $\mathcal{P}$ is broken under spiral spin configuration.

Table I. $\ell_2$ and winding angle ($\theta_0$ for proper screw and $\varphi_0$ for cycloid) dependent MPG and symmetrically allowed NSC and MIC components for wavevector $\boldsymbol{q} = \left(\frac{2n_2}{\ell_2}, \frac{n_2}{\ell_2}\right)$. Symbol ✗ indicates that all components are forbidden. We denote special angle series $\gamma_p = \left(p + \frac{p' n_2}{\ell_2}\right)\pi$ and $\gamma_{\frac{2p+1}{2}} = \left(\frac{2p+1}{2} + \frac{p' n_2}{\ell_2}\right)\pi$. Here $p = 0,1$ and $p' \in \mathbb{Z}$. The incident light is linearly polarized along principal axes $x$ and $y$.

| $\ell_2$ | Winding angles | MPG | NSC | MIC |
|---|---|---|---|---|
| Odd | $\theta_0 = \gamma_p$ | 2'2'2 | ✗ | ✗ |
| | $\theta_0 = \gamma_{\frac{2p+1}{2}}$ | 2'22' | ✗ | $\eta_{\text{prop}}^{yxx}, \eta_{\text{prop}}^{yyy}$ |
| | Other $\theta_0$ | 2' | $\sigma_{\text{prop}}^{xxx}, \sigma_{\text{prop}}^{xyy}$ | $\eta_{\text{prop}}^{yxx}, \eta_{\text{prop}}^{yyy}$ |
| | $\varphi_0 = \gamma_p$ | m'm2' | $\sigma_{\text{cyc}}^{yxx}, \sigma_{\text{cyc}}^{yyy}$ | ✗ |
| | $\varphi_0 = \gamma_{\frac{2p+1}{2}}$ | m'm'2 | $\sigma_{\text{cyc}}^{yxx}, \sigma_{\text{cyc}}^{yyy}$ | ✗ |
| | Other $\varphi_0$ | m' | $\sigma_{\text{cyc}}^{yxx}, \sigma_{\text{cyc}}^{yyy}$ | ✗ |
| Even | $\theta_0 = \gamma_p$ or $\theta_0 = \gamma_{\frac{2p+1}{2}}$ | 222.1' | ✗ | ✗ |
| | Other $\theta_0$ | 2.1' | $\sigma_{\text{prop}}^{xxx}, \sigma_{\text{prop}}^{xyy}$ | ✗ |
| | $\varphi_0 = \gamma_p$ or $\varphi_0 = \gamma_{\frac{2p+1}{2}}$ | mm2.1' | $\sigma_{\text{cyc}}^{yxx}, \sigma_{\text{cyc}}^{yyy}$ | ✗ |
| | Other $\varphi_0$ | m.1' | $\sigma_{\text{cyc}}^{yxx}, \sigma_{\text{cyc}}^{yyy}$ | ✗ |

*Tight-binding model calculation.* To explicitly illustrate the directional BPV generation, we take a minimum showcase with $\boldsymbol{q} = \left(\frac{1}{2}, \frac{1}{4}\right)$ [Figs. 1(a) and 1(b)] to perform tight-binding model calculations [47,48]. Without loss of generality, the basis spinor functions are chosen to be $\left[d_{xy}^\uparrow, d_{x^2-y^2}^\uparrow, d_{z^2}^\uparrow, d_{xy}^\downarrow, d_{x^2-y^2}^\downarrow, d_{z^2}^\downarrow\right]$. The Hamiltonian reads

$$\widehat{H} = \sum_{i,\mu} c_{i\mu,\alpha}^\dagger (\boldsymbol{\sigma} \cdot \boldsymbol{M}_i)_{\alpha\beta} c_{i\mu,\beta} + \sum_{\langle i,j\rangle,\mu\nu,\alpha} t_{i\mu,j\nu} c_{i\mu,\alpha}^\dagger c_{j\nu,\alpha} + \sum_{i,\mu,\alpha} U_\mu c_{i\mu,\alpha}^\dagger c_{i\mu,\alpha}, \quad (4)$$

where $c_{i,\mu}^\dagger$ ($c_{i,\mu}$) is the creation (annihilation) operator for orbital $\mu$ at site $i$. $\boldsymbol{\sigma}$'s are Pauli matrices for spin degree of freedom, characterizing the helical spin pattern with $\boldsymbol{M}_i$ defined in Eq. (1). $t_{i\mu,j\nu}$ denotes the hopping integral, which can be expressed by the $d$–$d$ interactions ($V_{dd\sigma}$, $V_{dd\pi}$, and $V_{dd\delta}$) according to the Slater-Koster scheme. $U_\mu$ is the orbital on-site energy, denoting the chemical potential difference between the $d_{z^2}$ and ($d_{xy}, d_{x^2-y^2}$). We also include on-site intrinsic SOC as $\widehat{H}_{SOC} = t_{SOC}\sigma_z \otimes h$, with $t_{SOC}$ representing the SOC strength and $h = \begin{pmatrix} 0 & -2i & 0 \\ 2i & 0 & 0 \\ 0 & 0 & 0 \end{pmatrix}$.

The band structures are shown in Figs. 1(c) and 1(d), and their spin textures can be found in Figs. S3 and S4 [37]. In the proper screw configuration, we take a general initial polar angle of $\theta_0 = \frac{\pi}{18}$ (as $\theta_0 = 0$ totally forbids BPV, see Table I), and the $\mathcal{C}_{2x}\mathcal{T}$ constraints its electric polarization $P$ along $x$. For the cycloid pattern, we use $\varphi_0 = 0$, giving the SOC-induced $P$ along $y$. In both cases, reversing spiral chirality (from $q$ to $-q$) would flip $P$ to $-P$.

We now move to the BPV generation under LPL, which generates NSC and MIC in the form [18]

$$J^a_{NSC} = \sigma^{abb}(0; \omega, -\omega)E_b(\omega)E_b(-\omega), \tag{5}$$

$$J^a_{MIC} = \eta^{abb}(0; \omega, -\omega)E_b(\omega)E_b(-\omega). \tag{6}$$

Here, the $\sigma^{abb}$ and $\eta^{abb}$ are the photoconductivity tensors. $E$ is the light alternating electric field (with angular frequency $\omega$), and $a$ ($b$) refers to Cartesian coordinate in the $xy$ plane. The NSC is irrespective to $\mathcal{T}$, which can be evaluated by [3,49]

$$\sigma^{abb}(0; \omega, -\omega) = \frac{\pi e^3}{\hbar^2} \int \frac{d^3k}{(2\pi)^3} \sum_{m,n} f_{mn} R^{a;b}_{mn} |r^b_{mn}|^2 \delta(\omega_{mn} - \omega), \tag{7}$$

where $f_{mn}$ and $\hbar\omega_{mn}$ measure the occupation and eigenenergy differences between $m$ and $n$ states, respectively. The interband position operator is $r^b_{mn} = \frac{\langle m|v^b|n\rangle}{i\omega_{mn}}$, $m \neq n$ ($v^b$ is the velocity operator). The gauge-invariant shift vector is $R^{a;b}_{mn} = \partial_a \phi^b_{mn} - \mathcal{A}^a_{mm} + \mathcal{A}^a_{nn}$, with $\phi^b_{mn}$ being the phase of $r^b_{mn}$ ($= |r^b_{mn}|e^{i\phi^b_{mn}}$) and $\mathcal{A}^a_{mm} = i\langle m|\partial_a m\rangle$ the intraband Berry connection. Hence, the NSC evaluates the contribution of the shift vector weighted by absorption rate $|r^b_{mn}|^2 \delta(\omega_{mn} - \omega)$ at each $k$.

The $\mathcal{T}$-odd MIC characterizes the band velocity differences between the valence and conduction bands [18],

$$\eta^{abb}(0; \omega, -\omega) = -\frac{\tau \pi e^3}{\hbar^2} \int \frac{d^3k}{(2\pi)^3} f_{mn} \Delta^a_{mn} g^{bb}_{mn} \delta(\omega_{mn} - \omega). \tag{8}$$

Here, $\Delta^a_{mn} = v^a_{mm} - v^a_{nn}$ denotes the velocity difference and the quantum metric tensor $g^{bb}_{mn} = |r^b_{mn}|^2$ is the real and symmetric part of the Bloch wavefunction. The MIC linearly grows with time and saturates at carrier relaxation time $\tau$, which is multiplied as a pre-factor in Eq. (8).

Our numerical calculations give consistent results with previous group theory. Taking the proper screw pattern as an example, one always has $\mathcal{C}_{2x}\mathcal{T} R^{y;b}_{mn}(k_x, k_y) = -R^{y;b}_{mn}(-k_x, k_y)$ ($b = x, y$). Hence, $\sigma^{yyy}$ and $\sigma^{yxx}$ are symmetrically forbidden, leaving only $\sigma^{xxx}$ and $\sigma^{xyy}$ to be finite [Fig.

2(a)]. As the $C_{3z}$ is broken under the helimagnetic pattern, these two components are independent of each other. This demonstrates that the NSC is unidirectional under proper screw configuration. Note that the trigonal lattice cannot host NSC generation under nonmagnetic, ferromagnetic, or collinear antiferromagnetic configurations. We illustrate the $\theta_0$-dependent NSC generation in Fig. S5 [37]. It shows that when $\theta_0 = 0, \frac{\pi}{4}, \frac{\pi}{2}, \frac{3\pi}{4}, \pi, ...$, the NSC is forbidden, in accordance with Table I. If one fix incident photon energy and track the NSC as a function of $\theta_0$, it shows a sinusoidal shape with nodal points at these angle values. It indicates that in addition to $\boldsymbol{q} \to -\boldsymbol{q}$ (Fig. S6 [37]) that could flip NSC direction, $\theta_0$ also controls the NSC magnitude and direction.

For the cycloid pattern, its $x$-propagation current will be forbidden as $C_{2y} R_{nm}^{x;b}(k_x, k_y) = -R_{nm}^{x;b}(-k_x, k_y)$, so that the NSC is also unidirectional and only transports along $y$, as numerically calculated in Fig. 2(b). This is in accordance with the fact that $\boldsymbol{P}$ along $y$. As for the MIC, it is found that they are all forbidden, since the $\mathcal{T}\boldsymbol{t}$ (time-reversal multiplies lattice translation) is always preserved in $\boldsymbol{q} = \left(\frac{1}{2}, \frac{1}{4}\right)$, see Table I.

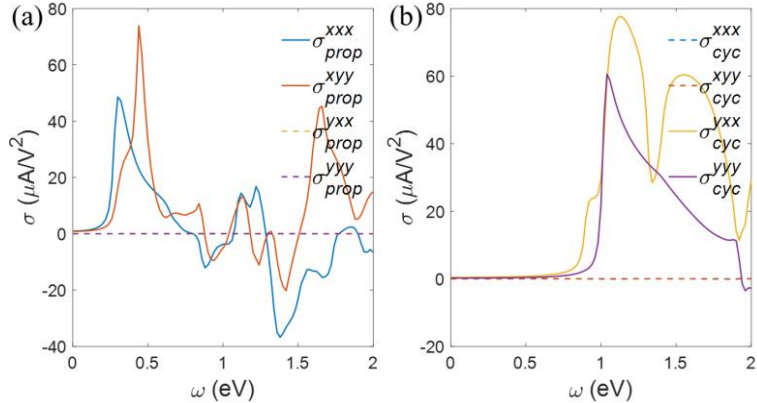

FIG. 2. Tight-binding calculated NSC photoconductance for $\boldsymbol{q} = \left(\frac{1}{2}, \frac{1}{4}\right)$ in (a) the proper screw ($\theta_0 = \frac{\pi}{18}$) and (b) the cycloid ($\varphi_0 = 0$) configurations. We assume an artificial layer thickness of 0.5 Å.

One may wonder how to control the BPV magnitude. In order to see this, we adjust the tight-binding parameters (Fig. S7) [37]. One clearly sees that smaller bandgap usually corresponds to larger NSC peak at the band edge, in accordance with larger light absorbance. Additionally, note that NSC also scales with the shift vector which reflects symmetry constraints. We find that in general a long helimagnetic wavelength (small $\boldsymbol{q}$) yields the system towards the collinear ferromagnetic pattern, which has simultaneous smaller bandgap and more symmetric shift vector. The competing of these two factors determines the ultimate NSC photoconductance magnitude. We find that $\boldsymbol{q} = \left(\frac{1}{4}, \frac{1}{8}\right)$ exhibits comparable magnitude with $\boldsymbol{q} = \left(\frac{1}{2}, \frac{1}{4}\right)$ (Figs. S8 and S9 [37]). In addition, when we set $\boldsymbol{q} =$ $\left(\frac{2}{3}, \frac{1}{3}\right)$ (proper screw with $\theta_0 = \frac{\pi}{18}$), directional MIC emerges (Fig. S10 [37]), consistent with Table I for odd $\ell_1$ and $\ell_2$.

*Realistic materials.* We perform density functional theory (DFT) calculations [37] in monolayer $NiI_2$ as a showcase. As plotted in Fig. 3(a), the Ni sites exhibit a trigonal lattice, and the iodine atoms form two atomic layers [50]. It belongs to a crystalline layer group of $P\bar{3}m1$. Experimentally, the monolayer $NiI_2$ can be exfoliated from its bulk counterpart and is characterized in the proper screw and cycloidal spin helix, depending on its substrates [33-36]. The thin film $NiI_2$ has been receiving hectic attention due to its unique helical spin order and the potential multiferroic order very recently [25,51-54].

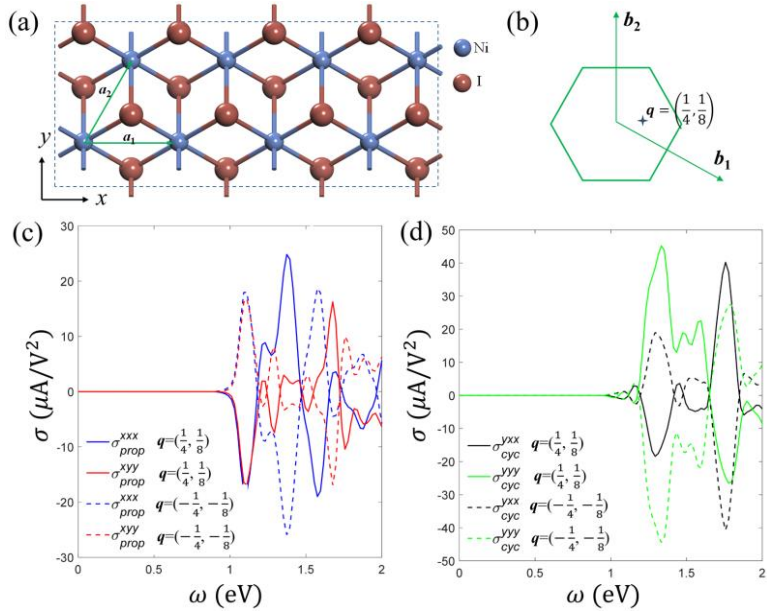

FIG. 3. (a) Top view of monolayer $NiI_2$ with the dashed rectangle denotes simulation magnetic supercell. (b) The first BZ of trigonal lattice and the position of $q$. NSC photoconductances under (c) proper screw and (d) cycloid spiral states.

Previous works suggest that the monolayer $NiI_2$ prefers a $q = \pm \left(\frac{1}{4}, \frac{1}{8}\right)$ helical spin [27,55,56] [Fig. 3(b)], and we will use this pattern in both proper screw and cycloid patterns. Our DFT calculations suggest that the proper screw (with $\theta_0 = 0$) is energetically lower than the cycloid spiral state ($\varphi_0 = 0$) by 3 meV per formula unit. No matter which specific pattern is chosen, the calculated bandgap is ~0.9 eV, and their band structures are plotted in Fig. S11 [37]. We calculate their electric polarization, giving $\mathbf{P} = (-2.85 \times 10^{-12}, 0, 0)$ C/m for the proper screw pattern, and $\mathbf{P} = (0, -9.67 \times 10^{-13}, 0)$ C/m for the cycloid configuration. All these results agree well with the symmetry consideration and previous results [33,36].

We plot the DFT calculated BPV photoconductance in Figs. 3(c)–3(d) and Figs. S12–S13 [37]. For the proper screw pattern, its NSC flowing along *x*. One sees that $\sigma_{\text{prop}}^{xxx}$ reaches 20 μA/V$^2$ if we take an effective thickness of 3 Å. Note that this current only arises when the spin spiral emerges, and it depends on the spiral patterns. Compared with the NSC values in nonmagnetic polar systems [57,58], this is large enough to be observed. As for the cycloid spiral state, the NSC is unidirectionally along *y*. The $\sigma_{\text{cyc}}^{yyy}$ value is approaching 50 μA/V$^2$ and $\sigma_{\text{cyc}}^{yxx}$ reaches 40 μA/V$^2$.

*Discussion and conclusion.* We would like to make a few remarks. Firstly, here we mainly focus on the symmetry arguments on how spin spiral controls magnetic group and BPV. Hence, the DFT calculations do not involve quasiparticle or exciton effect, which would red-shift the BPV peak position but is computationally demanding. According to previous works [59], many body effect does not significantly alter the NSC photoconductance owing to the ultrathin nature of 2D monolayers, and the DFT results can be comparable with experimental observations. Secondly, if the system lacks of SOC, the spin and lattice will be decoupled, and it would be the spin group theory to determine BPV generations. In such situations, all BPV components vanish. Thirdly, the monolayer NiI$_2$ is an exemplary platform for helimagnetic materials, and there are various similar systems that could be applied in this theory. We propose that the direction-dependent BPV generation could serve as a potential tool to identify their symmetry and ferroic orders, in addition to neutron diffraction, spin-resolved scanning tunneling microscope, or linear dichroism. Fourthly, other processes such as nonlinear Hall effect and second harmonic generation also arise from Berry curvature dipole and quantum metric dipole, and follow similar symmetry arguments [60-63]. Our results can be directly applied to explore their emergence in helimagnetic systems. Lastly, here we only evaluate the regular shift current generation, while the recently proposed Umklapp shift current in centrosymmetric charge density wave phase [64] is not considered.

In conclusion, we use magnetic group theory to show how the helimagnetic pattern controls the nonlinear optical bulk photovoltaic effect, which expands the previous discussions on nonmagnetic and antiferromagnetic systems. The unidirectional shift current and injection current generations depend on magnetic helicity, wavevector, and the specific winding angle. We perform tight binding and DFT calculations to verify the theory, which may serve as a complementary detection scheme for multiferroic materials and magnon excitations, in addition to other widely-used experimental efforts such as neutron diffraction, second harmonic generation, spin-resolved scanning tunneling microscope, or linear dichroism.

**Acknowledgments.** C.Z. thanks valuable discussions with Prof. Hongming Weng and Ziyin Song at Chinese Academy of Sciences. We acknowledge the financial support from the National Natural Science Foundation of China (NSFC) under Grant Nos. 12274342 and 12374065. Additionally, the authors acknowledge support from Beijing PARATERA Technology Co., LTD for providing high-performance resources for contributing to the research results reported within this paper.